\documentclass[aip,cha,preprint,numerical,onecolumn]{revtex4-1}
\usepackage{graphicx}
\usepackage{amsmath}
\usepackage{amssymb}
\usepackage[utf8]{inputenc}
\usepackage[hidelinks]{hyperref}
\usepackage{epstopdf}

\newcommand{\loc}[1]{\bar#1}
\newcommand{\im}[1]{\mbox{Im}\left[ #1 \right]}
\newcommand{\mean}[1]{\langle #1 \rangle}
\newcommand{\abs}[1]{\left\vert #1 \right\vert}

\def \t  {\theta}
\def \s  {\sigma}
\def \w  {\omega}
\def \W  {\Omega}

\def \g  {\gamma}
\def \vp {\varphi}
\def \p  {\psi}
\def \T  {\Theta}
\def \P  {\Phi}

\graphicspath{{./figures/}}

\begin{document}

\title{Partial synchronization in the Kuramoto model with attractive and
repulsive interactions via the Bellerophon state}
\date{\today}
\author{Erik Teichmann}
\email{kontakt.teichmann@gmail.com}
\affiliation{Institute of Physics and Astronomy, University of Potsdam,
Karl-Liebknecht-Str. 24/25, 14476 Potsdam-Golm, Germany}
\author{Rene O. Medrano-T}
\affiliation{Departamento de F\'isica, Universidade Federal de S\~ao Paulo,
Campus Diadema, R. S\~ao Nicolau, 210, 09913-030 SP, Brazil and
Departamento de F\'isica, Universidade Estadual  Paulista, Instituto de
Geoci\^encias e Ci\^encias Exatas, Campus Rio Claro, Av. 24A, 1515, 13506-900 SP,
Brazil}

\begin{abstract}
We study two groups of nonidentical Kuramoto oscillators with differing
  frequency distributions. Coupling between the groups is repulsive, while
  coupling between oscillators of the same group is attractive.  This asymmetry
  of interactions leads to an interesting synchronization behavior. For small
  coupling strength, the mean-fields of both groups resemble a more weakly
  coupled Kuramoto model. After increasing the coupling strength beyond a
  threshold, they reach the Bellerophon state of multiple clusters of averagely
  entrained oscillators. A further increase in the coupling strength then
  asymptotically transitions from the Bellerophon state to a single synchronized
  cluster. During this transition, the order parameters of both groups increase
  and resemble an equally strongly coupled Kuramoto model. Our analysis is based
  on the Ott-Antonsen mean-field theory.
\end{abstract}


\maketitle

\section{Introduction}

Synchronization describes the emergence of a common mode in a system of
interacting units. It has been observed in many different contexts, such as
power grids~\cite{motter2013spontaneous}, neuronal
populations~\cite{uhlhaas2009neural}, or the synchronization of pedestrians on a
bridge~\cite{strogatz2005theoretical}. One of the most popular models in the
study of synchronization is the Kuramoto
model~\cite{kuramoto1984chemical,acebron2005kuramoto,pikovsky2015dynamics}. It
is widely used for its description of the transition to synchronization and its
analytical
solvability~\cite{watanabe1993integrability,watanabe1994constants,ott2008low,ott2009long,mihara2019stability}.
In the model, only attractive pairwise interactions are considered. However,
there exist many systems that contain both attractive and repulsive
interactions~\cite{peyrache2012spatiotemporal}.

A method to model these mixed-interaction systems is the splitting of the
oscillator population into two groups. In a conformists-contrarian model, the
conformists interact attractively with all oscillators, and the contrarians
repulsively~\cite{hong2011conformists,hong2011kuramoto,hong2014periodic,qiu2016synchronization}.
One possible state in such a system with non-identical oscillators is the
Bellerophon state~\cite{qiu2016synchronization}. In this state, clusters exist,
where oscillators share the same average frequency but not the same phase or
instantaneous frequencies. The clusters have a constant difference in their
average frequency, yielding a step-like picture. Oscillators between the
clusters behave quasiperiodically.

Each oscillator in such a model of conformists and contrarians has one type of
interaction. Another possibility is to differentiate between groups in a model
with attractive and repulsive
interactions~\cite{montbrio2004synchronization,sheeba2008routes,anderson2012multiscale,hong2012mean,laing2012disorder,lohe2014conformistcontrarian,sonnenschein2015collective,pietras2016equivalence,kotwal2017connecting,achterhof2020two,achterhof2020twoa}.
Typically in this case, each oscillator acts attractively on all oscillators
of the same group and repulsively on oscillators of the other group.  This model
is a very simplified representation of a two-party dynamic, where people in the
same party try to agree but at the same time distance themselves from the other
party.

This paper considers such a model with attractive and repulsive interaction with
independent frequency distributions of the oscillators. For small coupling
strength, the macroscopic dynamics of each group resemble the ones of a Kuramoto
model with a weaker coupling. The macroscopic frequencies of both groups are
entrained, and there exists one entrained cluster. An increase in the coupling
strength destroys the entrainment, and the oscillators begin to form loose
clusters with a constant average frequency, separated by quasiperiodic ones;
they reach a Bellerophon state. A further increase in the coupling strength
leads to a jump in the synchronization of each group and an entrainment of both
groups' macroscopic dynamics. The entrained clusters then still exist but start
to collapse and approach one big entrained cluster, like the one expected for
the Kuramoto model without groups. The observed Bellerophon state is very
susceptible to changes in the frequency distribution.

The paper is organized as follows. First, the model is introduced. The solution
for the mean-field in the thermodynamic limit is then found with the Ott-Antonsen
Ansatz. Using the Ott-Antonsen equations, the mean-field dynamics are
investigated for increasing coupling strength, and the behavior is compared to
the Kuramoto model. Finally, the findings are summarized.

\section{The model}

The standard Kuramoto model~\cite{kuramoto1984chemical,acebron2005kuramoto} only
considers a single group of nonidentical oscillators. To describe a system of
$M$ interacting groups, the M-Kuramoto
model~\cite{qiu2016synchronization,maistrenko2014solitary} is needed. It is
defined as
\begin{equation}
  \dot{\t}^{\s}_j = \w^{\s}_j + \sum_{\s' = 1}^M \frac{K_{\s\s'}}{N}
    \sum_{k = 1}^{N_{\s'}} \sin(\t^{\s'}_k - \t^{\s}_j) \; .
  \label{eq:m_kuramoto}
\end{equation}
Here $\w^{\s}_j$ is the natural frequency of the $j$th oscillator in group $\s$
and $K_{\s\s'}$ is the coupling strength between groups $\s$ and $\s'$. Every
group consists of $N_{\s}$ oscillators with a total system size of $N =
\sum_{\s'} N_{\s'}$. In a simple model with attractive and repulsive
interactions, there are only two groups. To further simplify this, we consider
equal intra-coupling and inter-coupling $K_{\s\s'} = 2K$ if $\s = \s'$ and
$K_{\s\s'} = -2K$ else, respectively, and equally sized groups $N_1 = N_2$. For
the sake of analytical tractability, the oscillators' natural frequencies are
chosen from a Lorentzian distribution $g(\w)$
\begin{equation}
  g(\w) = \frac{1}{\pi \g \left[1 + \left(\frac{\w - \loc{w}}{\g}\right)^2 \right]}  \; ,
  \label{eq:lorentz}
\end{equation}
where both groups have different centers $\loc{w}$ and widths $\g$. By rotating
the reference frame, the first group can be centered at $\loc{\w_1} = 0$, and by
scaling of the time, the width of the first group can be set to $\g_1 = 0.1$.
Without loss of generality, the first group is always chosen to be the narrower
one, i.e., $\g_2 \geq 0.1$. This model can also be represented as a single group
but with a bimodal frequency distribution and one additional bifurcation
parameter~\cite{pietras2016equivalence}.

Introducing the mean field for each group $Z_{\s} = R_{\s}e^{i\T_{\s}} =
(1/N_{\s}) \sum_k e^{i \t^{\s}_{k}}$ and renaming $\t^1_j = \vp_j$ and $\t^2_j =
\p_j$ yields the system
\begin{equation}
  \begin{aligned}
    \dot{\vp}_j = & \w_{1, j} + K R_1 \sin(\T_1 - \vp_j)
        - K R_2 \sin(\T_2 - \vp_j) \; , \\
    \dot{\p}_j = & \w_{2, j} - K R_1 \sin(\T_1 - \p_j)
        + K R_2 \sin(\T_2 - \p_j) \; .
    \label{eq:contrarian_mean_field}
  \end{aligned}
\end{equation}
Note that $K = K_{\s\s'}/2$, so the coupling strength of each group is halved
compared to the M-Kuramoto model. From there, the forcing $H$ for each group can
be determined to be
\begin{equation}
  H_{\s} = h_{\s} e^{i\P_\s} = K(Z_{\s} - Z_{\s'}) \; ,
  \label{eq:forcing}
\end{equation}
where $\s'$ denotes the other group. Both forcings are connected via $H_2 =
-H_1$, so their magnitudes are identical, $h_1 = h_2 = h$, and they are just
shifted by a phase $\P_1 = \P = \P_2 - \pi$.  The forced equations for the
oscillators are then
\begin{equation}
  \begin{aligned}
    \dot{\vp}_j = & \w_{1, j} + \im{H_1 e^{-i \vp_j}}
        = \w_{1, j} + h \sin(\P - \vp_j) \; , \\
    \dot{\p}_j = & \w_{2, j} + \im{H_2 e^{-i \p_j}}
        = \w_{2, j} + h \sin(\P - \p_j + \pi) \; .
    \label{eq:contrarian_forcing}
  \end{aligned}
\end{equation}

\subsection{Thermodynamic limit}

Systems consisting of the imaginary part of a forcing, like
Eqs.~\eqref{eq:contrarian_forcing} separately, have an analytical description of
their mean-field in the thermodynamic limit. The mean-field evolution is found
with the Ott-Antonsen (OA)
Ansatz~\cite{ott2008low,ott2009long,pikovsky2008partially} and in the case of a
Lorentzian distribution of the frequencies $\w$ the mean-field dynamics in the
OA Ansatz are ~\cite{pazo2020winfree}
\begin{equation}
  \dot{Z} = (- \g + i\loc{\w}) Z + \frac{1}{2} (H - H^*Z^2) \; ,
\end{equation}
with $H^*$ being the complex conjugate of $H$. By using the definition of the
forcing in Eq.~\eqref{eq:forcing} this becomes
\begin{equation}
  \dot{Z}_{\s} = (-\g_{\s} + i\loc{\w_{\s}}) Z_{\s} + \frac{K}{2}
    (Z_{\s}(1 - R_{\s}^2 + Z^*_{\s'}Z_{\s}) - Z_{\s'}) \; .
    \label{eq:oa_full_mean_field}
\end{equation}
Calculating the derivative of the mean-field $\dot{Z}_{\s} = \dot{R}_{\s} e^{i
\T_{\s}} + i R_{\s} \dot{\T}_{\s} e^{i\T_{\s}}$, multiplying
Eq.~\eqref{eq:oa_full_mean_field} by $e^{-i \T_{\s}}$ and splitting the real and
imaginary part yields the dynamics of the order parameter and the phase of the
mean-field as
\begin{equation}
  \begin{aligned}
    \dot{R}_{\s} = & - \g_{\s}R_{\s}
        + \frac{K}{2} (1 - R_{\s}^2)(R_{\s} - R_{\s'}\cos(\T_{\s} - \T_{\s'}))
        \; , \\
    \dot{\T}_{\s} = & \loc{\w_{\s}}
        + \frac{K}{2} \frac{R_{\s'}}{R_{\s}} (1 + R_{\s}^2)
        \sin(\T_{\s} - \T_{\s'}) \; .
  \end{aligned}
  \label{eq:oa_mean_field_group}
\end{equation}
Both equations only depend on the difference between the mean-field phases, as
expected in a rotational invariant system. The 4-dimensional system (order
parameter and mean-field phase of each group) can then be reduced to a
3-dimensional system of the order parameters and the phase difference $\T = \T_1
- \T_2$ to
\begin{equation}
  \begin{aligned}
    \dot{R}_{1} = & - \g_{1}R_{1}
        + \frac{K}{2} (1 - R_{1}^2)(R_{1} - R_{2}\cos\T)
        \; , \\
    \dot{R}_{2} = & - \g_{2}R_{2}
        + \frac{K}{2} (1 - R_{2}^2)(R_{2} - R_{1}\cos\T)
        \; , \\
    \dot{\T} = & - \loc{\w_{2}}
        + \frac{K}{2} \sin\T \frac{R_1^2(R_2^2 + 1) + R_2^2(R_1^2 + 1)}{R_1R_2}
        \; .
  \end{aligned}
  \label{eq:mean_field_reduced}
\end{equation}

Indicator oscillators can be used to study the system in the thermodynamic
limit. They are coupled to the OA mean-fields from
Eqs.~\eqref{eq:oa_mean_field_group} but do not create a mean-field themselves.
From Eqs.~\eqref{eq:contrarian_forcing}, the same forcing acts on both groups, so
their dynamics will be equal for the same $\w_j$ up to the phase shift. The
dynamics of a single indicator oscillator depends only on the oscillator's
frequency, not on the group it belongs to.  To discern the phases of indicator
oscillators, they are marked as $\t'_{j}$ with their natural frequencies
$\w'_j$. They follow the equations (we choose to consider them as part of group
2)
\begin{equation}
  \dot{\t}'_j = \w'_{j} - K R_1 \sin(\T_1 - \t'_j)
      + K R_2 \sin(\T_2 - \t'_j) \; .
  \label{eq:indicator}
\end{equation}
The $R_{1,2}$ and $\T_{1,2}$ are calculated from
Eq.~\eqref{eq:oa_mean_field_group}, and the $\w'$ are chosen uniformly. This
method makes it possible to calculate the observed frequencies $\dot{\t}'$ and
the phase distribution in the thermodynamic limit numerically.

\section{Small coupling strength}

In the Kuramoto model, i.e., Eq.~\eqref{eq:m_kuramoto} with $M = 1$, the
critical coupling for the emergence of a nonzero order parameter and the
function $\hat{R}(\hat{K})$ is known analytically in the case of a Lorentzian
distribution of the phases~\cite{acebron2005kuramoto}. The critical coupling for
the transition from incoherence to partial synchrony is
\begin{equation}
  \hat{K}_c = 2 \hat{\g} \; ,
  \label{eq:kuramoto_critical}
\end{equation}
and the order parameter
\begin{equation}
  \hat{R} = \sqrt{1 - \frac{\hat{K_c}}{\hat{K}}} \; .
  \label{eq:kuramoto_order_parameter}
\end{equation}
The hat notation ,$\hat{K}$, marks the quantities of the Kuramoto model.

In the considered model with $M = 2$, if both groups are identical, then the
critical coupling strength of each group corresponds to the Kuramoto model's
one~\cite{anderson2012multiscale}.  To compare the nontrivial order  parameter
between the Kuramoto model in Eq.~\eqref{eq:kuramoto_order_parameter} and the
model with attractive and repulsive interactions in
Eq.~\eqref{eq:oa_full_mean_field}, the coupling constants have to be scaled, as
$K_{\s\s'} = 2K$ was chosen and in the case of the Kuramoto model $K_{\s\s'} =
\hat{K}$, so $\hat{K} = 2K$. A comparable critical coupling $K_c$ for the
contrarian model would then correspond to $K_c = 2 \g = 4 \hat{\g}$, i.e., $\g =
2 \hat{\g}$. The order parameter will have the same form, as $K_c/K$ from
Eq.~\eqref{eq:kuramoto_order_parameter} only depends on $K_c$.

\begin{figure}
  \includegraphics{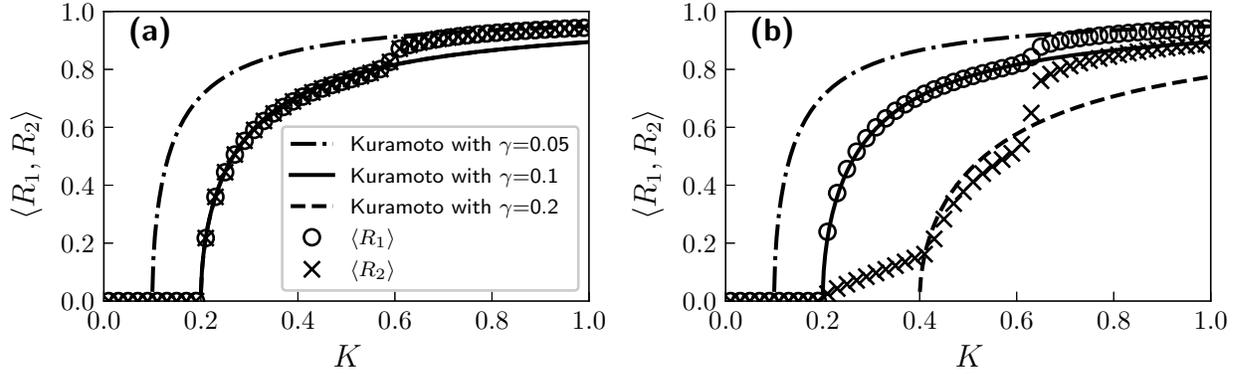}
  \caption{Average order parameters of group 1 $\mean{R_1}$ as circles and group
  2 $\mean{R_2}$ as crosses. The solid (dashed, dash-dotted) black line shows
  the average order parameter in a Kuramoto model with $\g = 0.1$ ($\g = 0.2$,
  $\g = 0.05$). In (a) $\g_2 = 0.1$, so both distributions have the same width
  but different centers, and in (b) $\g_2 = 0.2$. The width is $\loc{\w_2} =
  1$ in both cases.}
  \label{fig:order_parameters}
\end{figure}

\begin{figure}
  \includegraphics{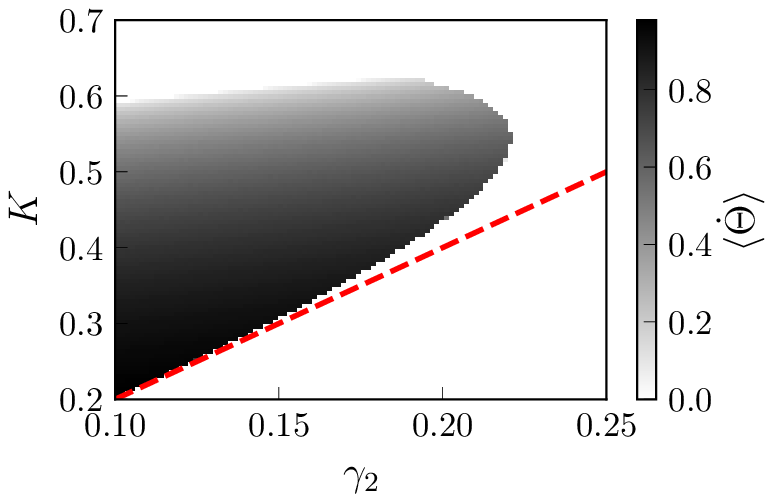}
  \caption{The difference in average frequencies of the mean fields
  $\mean{\dot{\T}}$ for $\loc{\w_2} = 1$. The red dashed line is the critical
  coupling of a Kuramoto model with the same width of the frequency distribution
  $\hat{\g} = \g$ as the second group in Eq.~\eqref{eq:kuramoto_critical}. The
  critical coupling for group 1 is $K = 0.2$, below this the system is
  incoherent ($R_1 = R_2 = 0$). For any scaling of $\loc{\w_2}$, this image
  looks identical, if $K$ and $\g_2$ scaled with the same factor (only the
  absolute values of the frequency difference change).}
  \label{fig:frequency_difference}
\end{figure}

For small coupling strength, both groups are incoherent. In
Fig.~\ref{fig:order_parameters}, it is shown that the critical coupling strength
for the first group corresponds to $K_{c,1} = 0.2$, the same as in a Kuramoto
model with $\hat{\g}_1 = \g_1 = 0.1$, not the expected $\hat{\g}_1 = \g_1/2$.
The effective coupling strength is only half as strong as in the Kuramoto model.
For $K \lessapprox 0.6$, the first group follows the relation in
Eq.~\eqref{eq:kuramoto_order_parameter} with $K_{c, 1} = 0.2$ very well,
regardless of $\g_2$, as can be observed from
Fig.~\ref{fig:order_parameters}(a) and (b), when $\g_2$ changes from
$0.1$ to $0.2$.  The second group behaves somewhat differently. Its critical
coupling follows a perturbed form of $K_{c, 2} = 2 \g_2$, half as strong as in
the Kuramoto model.  For $K_{c, 1} < K \lessapprox K_{c, 2}$ this group is not
incoherent but becomes forced by group 1, as shown in
Fig.~\ref{fig:order_parameters}(b).

Similar to the Kuramoto model, there
exists a cluster of averagely entrained oscillators, centered around $\w_j =
0$~\footnote{The center of group 1.} with a small shift in the direction of
$\loc{\w_2}$. They are entrained to the mean-field of group 1, and, as before,
the observed frequency only depends on the oscillator's natural frequency
$\w_j$, not on the group it belongs to~\footnote{Oscillators of group 2 are
shifted by $\pi$ in reference to oscillators of group 1 in this case.}. The
difference in the groups' mean-field frequencies can be seen in
Fig.~\ref{fig:frequency_difference}.  The mean-fields are entrained in their
average and instantaneous frequency since group 1 forces the, not yet
self-synchronized, group 2.

\section{Moderate coupling strength}

\begin{figure}
  \includegraphics{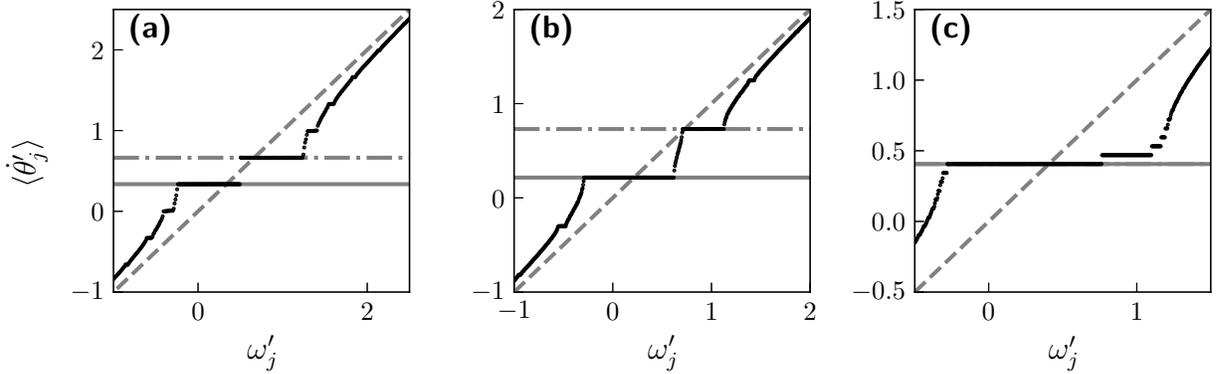}
  \caption{The average observed frequencies of indicator oscillators from
  Eqs.~\eqref{eq:oa_mean_field_group} and \eqref{eq:indicator}. The solid
  (dash-dotted) gray line marks the average mean field frequency
  $\mean{\dot{\T}_1}$ ($\mean{\dot{\T}_2}$) and the dashed gray line the
  identity function $\mean{\dot{\t}'_j} = \w'_j$. For all three subfigures
  $\loc{\w_2} = 1$. In (a) $\g_2 = 0.1$, $K = 0.56$ and identical order
  parameters $R_1 = R_2$, in (b) $\g_2 = 0.2$, $K = 0.56$ and $R_1 \neq R_2$,
  and in (c) $\g = 0.2$, $K = 0.63$ and $R_1 \neq R_2$ but the mean field
  frequencies are entrained.}
  \label{fig:average_frequencies}
\end{figure}

With the increase of the coupling strength to $K > K_{c, 2}$, new clusters begin
to form. These new clusters lead to a stronger synchronization of group 2, and
its order parameter better follows the relation
Eq.~\eqref{eq:kuramoto_order_parameter} but is smaller than expected, as group 1
perturbs it. The clusters are coherent regions in $\w$ with a common average
frequency, as shown in Fig.~\ref{fig:average_frequencies}. Oscillators between
the clusters behave quasiperiodically. The entrained clusters decrease in size
the further they are from the mean of the two distributions
$\loc{\w_2}/2$~\footnote{Note that $\loc{\w_1}$ has been chosen to be $0$.}, and
the two biggest clusters lock to the average frequency of the mean fields (if
they are non-entrained). Such a state has been observed in a
conformists-contrarian model with a common distribution of frequencies in
Ref.~\cite{qiu2016synchronization} where they called it the Bellerophon state,
noting the difference to the chimera state. There they determined the
entrainment frequencies to be uneven multiples of a fundamental frequency. This
entrainment to uneven multiples is also the case here, albeit the zero is
shifted. In the case of non-entrainment of the mean-fields, the clusters have an
average frequency of uneven multiples of the difference in average mean-field
frequencies $\Delta\W = \mean{\dot{\T}}/2$ centered around their average mean
$\W_0 = (\mean{\T_1} + \mean{\T_2}) / 2$, i.e., the frequencies of the clusters
$\W_j$ are
\begin{equation}
  \W_{\pm(2n+1)} = \W_0 \pm (2n+1)\Delta\W \; .
\end{equation}
The center frequency $\W_0$ is not constant but changes with the coupling
strength $K$ and $\g_2$.

The transition between the clusters with $\W_{-1}$ and $\W_{1}$ depends on
$\g_2$. If both widths $\g_1 = \g_2$ are identical, then there is a sudden
transition between the states (Fig.~\ref{fig:average_frequencies}(a)) while for
$\g_1 \neq \g_2$ the transition is
smooth(Fig.~\ref{fig:average_frequencies}(b)).  A similar state has also been
observed in Ref.\cite{montbrio2004synchronization} for a system of attractive
and repulsive oscillators with frequency distributions of common width but
different means.  The dynamics observed here differ from both, as
Ref.~\cite{qiu2016synchronization} only considers a conformists-contrarian model
and finds only the special case depicted in Fig.\ref{fig:average_frequencies}(a)
shows a sudden jump between the clusters with frequencies $\W_1$ and $\W_{-1}$.
The difference to Ref.~\cite{montbrio2004synchronization} lies in the perfect
overlap of the average frequencies of oscillators with the same $\w_j$,
regardless of their group and the different width of the frequency
distributions.

\begin{figure}
  \includegraphics{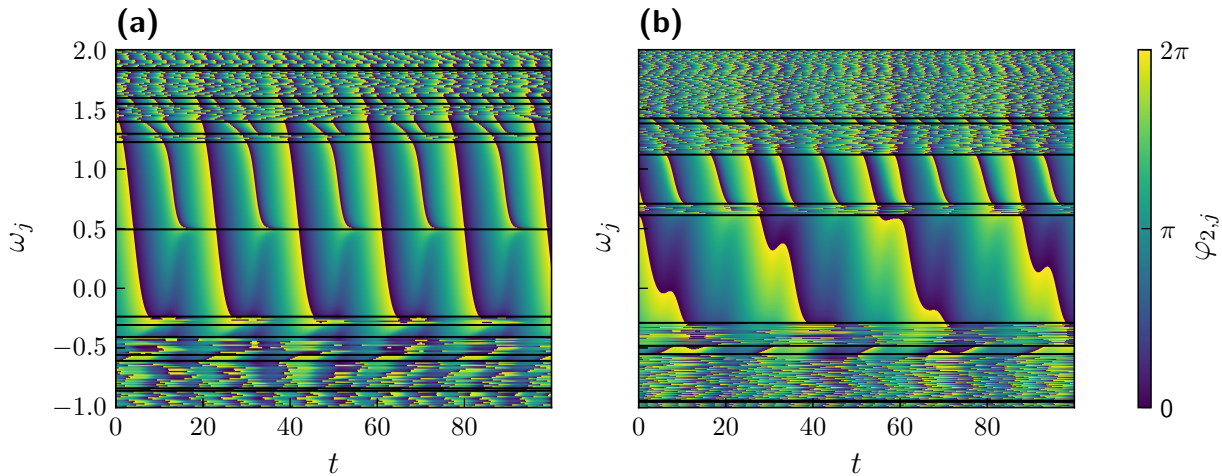}
  \caption{The phases of the indicator oscillators. Approximate cluster borders
  are marked with black lines. The parameters of (a) and (b) are the same as in
  Fig.~\ref{fig:average_frequencies} (a) and (b).}
  \label{fig:phase_evolution}
\end{figure}

The clusters are stable in time, as seen in Fig.~\ref{fig:phase_evolution}.
While the clusters are fully entrained, they do not share a common phase. The
difference in frequency can be seen very well by the additional rotation in the
cluster with $\W_1$ in reference to $\W_{-1}$.  Again, there is a difference in
the case of $\g_1 \neq \g_2$, where the dynamics have a more complicated phase
dynamic.

\begin{figure}
  \includegraphics{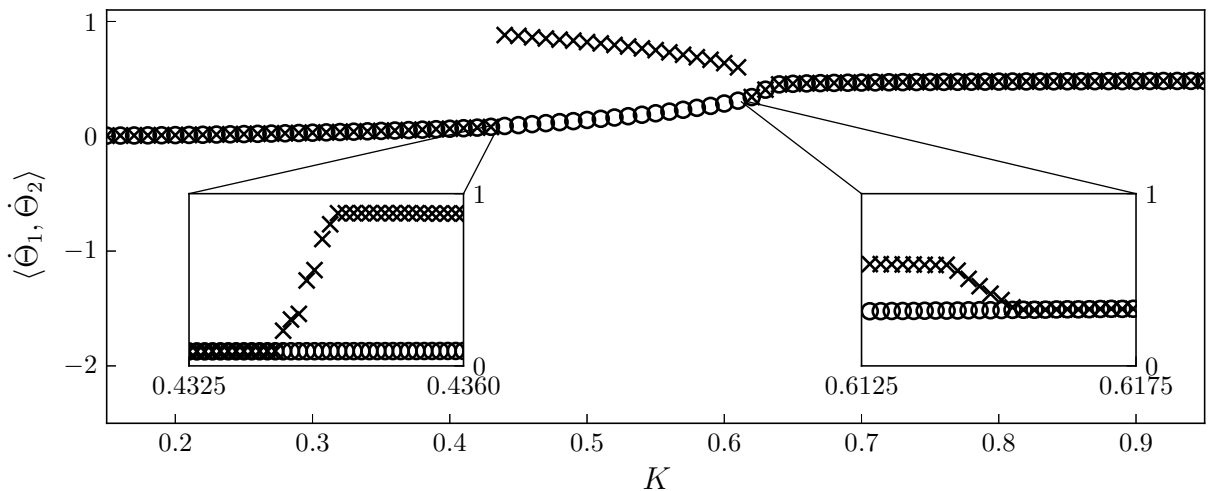}
  \caption{The average frequencies of the mean fields $\mean{\dot{\T}_1}$
  (circles) and $\mean{\dot{\T}_2}$ (crosses) for $\loc{\w_2} = 1$ and $\g_2 =
  0.2$. The inset figures show the transition from entrainment to
  non-entrainment and vice versa. Because the frequency has some very big peaks
  during the transition, all numerical values with $\abs{\dot{\T}_2} > 100$ were
  removed to smoothen the transition.}
  \label{fig:frequency_entrainment}
\end{figure}

The Bellerophon state comes into existence before the mean fields lose their
entrainment, but after $K_{c, 2}$. The clusters lead to the fast transition
between entrainment and non-entrainment of the mean-fields in
Fig.~\ref{fig:frequency_entrainment}.  In the case of $\g_1 = \g_2$, the forcing
$H$ has a unique role.  Its average frequency $\mean{\dot{\P}}$ lies precisely
between the clusters at $\W_0$. For $\g_1 \neq \g_2$ this is not the case, as
$\mean{\dot{\P}} = \mean{\dot{\T}_1} = \W_0 - \Delta\W$. The increase of the
difference in $K$ and $K_{c, 2}$ for non-entrainment with $\g_2$ is shown in
Fig.~\ref{fig:frequency_difference}.

\section{Strong coupling strength}

For strong coupling strength of $K \approx 0.6$, the mean-fields entrain again.
The clusters from the Bellerophon state persist but begin to shrink and approach
a single big cluster, as shown in Fig.~\ref{fig:average_frequencies}(c). The
mean-fields are perfectly entrained, not just in their average frequency, and
shifted by about $\pi$.  As soon as both mean-fields entrain, the mean-field of
group 2 adds a constant repulsive forcing to the oscillators of group 1, which
increases their coherence and leads to a higher order parameter. In the case of
a phase shift of $\pi$, both terms in the forcing in Eq.~\eqref{eq:forcing}
would be identical, and the oscillators are forced by $2K$, and resemble a
Kuramoto model with $\hat{K} = 2K$ (or a Kuramoto model with $\hat{K} = K$ and
$\g_1 = 2\hat{\g}_1$), as expected. The same effect also happens for group 2.
The phase shift between the two mean-fields is not $\pi$, so repulsion of the
two groups is not perfect, and the average order parameter only approaches the
Kuramoto model order parameter asymptotically. The closeness to the Kuramoto
model's order parameter depends on the difference in the width $\g_1$ and
$\g_2$. The closer they are (as in Fig.~\ref{fig:order_parameters}(a)), the
better the approximation will be, and the closer the phase shift will be to
$\pi$ for an equal $K$.

During the transition from non-entrainment to entrainment, the average
mean-field frequency of group 2 quickly drops, while the cluster with $\W_1$
persists and changes its position only slightly. This is a rather remarkable
observation as there no longer exists a second-mean field frequency to entrain
to. Instead, it shows that the middle-frequency $\W_0$ and the step-like
clusters are an intrinsic property of the system that is not simply generated
from the interaction of the mean fields. $\W_0$ is also different from the
middle of the two frequency distributions and depends explicitly on the coupling
strength. After a further increase in $K$, the distance between the clusters
reduces even further in Fig.~\ref{fig:average_frequencies}(c). With an increase
in $K$, the distance in $\w_j$ between the clusters decreases, and the jump
between $\W_1$ and $\W_3$ becomes discontinuous. A further increase in the
coupling strength finally asymptotically yields one entrained subpopulation,
like expected in the Kuramoto model for nonidentical oscillators. This again
verifies the connection between the M-Kuramoto model and the Kuramoto model
already seen in the order parameters.

\section{Numerical realization}

\begin{figure}
  \includegraphics{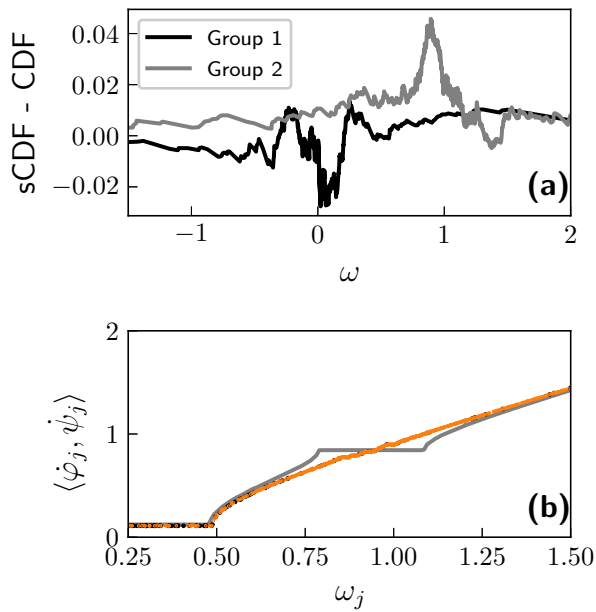}
  \caption{Numerical sampling leads to instabilities of the frequency
  distribution. This system uses the same parameters $\loc{\w_2} = 1$, $\g_2 =
  0.2$ and a similar coupling of $K=0.48$ to
  Fig.~\ref{fig:average_frequencies}(b) and Fig.~\ref{fig:frequency_entrainment}
  with $N_1 = N_2 = 1000$. In (a), the difference between the sampled cumulative
  distribution function (sCDF) of $\w_1$ and $\w_2$ and the cumulative
  distribution function (CDF) of the corresponding Lorentz distribution in
  Eq.~\eqref{eq:lorentz} is shown. In (b), the average frequencies are compared
  to the OA solution (gray line).  Group 1 is shown as black circles, and group
  2 as orange dots.}
  \label{fig:sampling_frequency}
\end{figure}

The Bellerophon state can be observed numerically for moderately big systems of
the size of, e.g., 800 or 1000 oscillators. Their mean-field dynamics and their
average frequencies fit very well to the predictions of the OA theory. In a few
cases, a significant difference can be observed for non-entrained
mean-fields\footnote{$1$ case of about $100$ observed systems with non-entrained
frequencies}.  One such case is shown in Fig.~\ref{fig:sampling_frequency}.
While the initial phase distribution is of not much importance, the sampling of
the frequencies is. In the shown case, the frequencies of group 2 were
oversampled close to $\W_3$, and group 1 was undersampled in $\W_1$. This
undersampling leads to a weaker order parameter $R_1$~\footnote{The clusters are
the main contributor to the order parameter}, which disturbs the oscillators
that would normally form the cluster $\W_3$ in
Fig.~\ref{fig:sampling_frequency}(b). This, in turn, leads to a significantly
weaker order parameter $R_2 \approx 0.22$ compared to the OA solution with $R_2
\approx 0.36$.

\section{Conclusion}

We have shown the three different partial synchronous states in a two-group
Kuramoto model with attractive and repulsive interaction. For weak coupling,
each group can be well described by a forced Kuramoto model with an equal
frequency distribution. For the group with the narrower distribution, the
forcing can be disregarded.

Shortly after the coupling strength exceeds both groups critical coupling
strength, there exists a Bellerophon state with multiple, step-like clusters of
averagely entrained oscillators. The oscillators' instantaneous dynamics in each
cluster differ. This state leads to a non-entrainment of both groups mean-field
frequencies. Still, they can be described by a similar Kuramoto model.

A further increase in the coupling strength leads to an entrainment, and the
Bellerophon state shrinks and asymptotically approaches a one cluster state, as
in the Kuramoto model. At this point, both groups synchronize stronger, and each
approaches a Kuramoto model with a frequency distribution half as wide as the
groups own.

The Bellerophon state is also reproduced numerically for moderately big systems
of about 1000 oscillators. In special cases of sampling errors in the frequency
distribution, the higher clusters are perturbed.  As a result, the order
parameter of group 2 becomes small compared to the analytical solution. So
special care has to be taken when realizing such a configuration.

Open problems are the calculation of the critical coupling for the increase in
synchronization and the order parameter of the second group in the Bellerophon
state, as it differs quite a bit from the Kuramoto model solution.

\section*{Data Availability}

The data that support the findings of this study are available from the
corresponding author upon reasonable request.

\begin{acknowledgements}
This paper was developed within the scope of the IRTG 1740 / TRP 2015/50122-0,
funded by the DFG/ FAPESP.
\end{acknowledgements}

\bibliography{lit}

\end{document}